\documentstyle[12pt]{article}

\def\pa{\partial}

\def\na{\nabla}
\def\nn{\nonumber}

\def\sg{\sqrt{-g}}
\def\ga{\gamma}

\def\si{\sigma}

\def\ka{\kappa}
\def\al{\alpha}

\def\la{\lambda}
\def\La{{\Lambda}}
\def\vphi{{\varphi}}

\def\const{{\rm const}}

\def\d{{\rm d}}
\def\tg{\tilde{g}}

\begin{document}
\thispagestyle{empty}
\begin{flushright}
hep-th/0305170
\end{flushright}
\vspace{0.5cm}
\begin{center}
{\Large \bf General warped solutions in 5D dilaton gravity}
\vspace{1cm}

Krzysztof A. Meissner\footnote{Krzysztof.Meissner@fuw.edu.pl}, 
Marek Olechowski\footnote{Marek.Olechowski@fuw.edu.pl}\\ 
\vspace{5mm}
{\it Institute of Theoretical Physics, Warsaw University\\ Ho\.za 69,
00-681 Warsaw, Poland}\\
\end{center}
\vspace{1cm}
\begin{center}
{\bf Abstract}
\end{center}

We present explicit analytic form of general warped solutions of the
string inspired dilaton gravity system with bulk cosmological constant 
in 5 dimensions. The general solution allows for either nonvanishing
effective 4--dimensional cosmological constant or the nontrivial
4--dimensional dilaton but not both.

\section{Introduction}

String theory is the leading candidate for the generalization of 
gravity. Indeed one of the massless string modes is the graviton 
and the effective action involves the Einstein--Hilbert action.
In the low energy string effective action there is necessarily also 
another field called the dilaton that couples to (almost) all other
fields. In this paper we consider the string-inspired dilaton
gravity system in 5 dimensions. We seek solutions of the  warped form
\cite{RSh,RS2} for the metric and as sum of  $x^\mu$ and  $y$
dependent parts for the dilaton (as usually we denote by $y$ the last
coordinate). Surprisingly the general solutions of the bulk equations
of motion can be found and we give their explicit analytic form in the
string frame. These solutions are the building blocks which, when
supplemented by appropriate branes, can be used to construct global
warped solutions.

The transition from the string frame 
to the usually used Einstein frame is immediate but the results in
general cannot be given in the explicit analytic form -- we present in
the paper such a transition for the cases where it can be explicitly
done. Some special solutions for this system in the Einstein frame
were obtained previously in \cite{ADKS,KSS1,KSS2,AJS,FKV,NO} (we do not
consider here theories with other bulk scalar fields,
often also called dilaton or radion fields, which may have different
form of interactions).

\section{5--dimensional dilaton gravity}

The 5--dimensional gravity--dilaton action we consider is the tree level 
string effective action and reads in the string frame:
\begin{equation}
S=\frac1{2\ka_5^2}\int \d^4 x dy
\sg e^{-\vphi}\left(-2\La+R^{(5)}+\pa_\mu\vphi\pa^\mu\vphi\right).
\label{action}
\end{equation}
The equations of motion for the action (\ref{action}) read
\begin{eqnarray}
0 &=&
R_{\mu\nu}^{(5)}+\na_\mu\pa_\nu\vphi
\,,
\label{T0}
\\[2pt]
0&=& 2\La +\pa_\mu\vphi\pa^\mu\vphi -\Box^{(5)}\vphi
\,.
\label{W0}
\end{eqnarray}
We are looking for solutions for which the metric has the warped form
\begin{equation}
ds^2=e^{2A(y)} g_{\al\beta}(x) dx^\al dx^\beta+dy^2
\,
\label{ds2}
\end{equation}
and the dilaton field separates
\begin{equation}
\vphi(x^\al,y)=\si(x^\al)+\phi(y)\,,
\end{equation}
where the indices $\al,\beta=0,\ldots,3$ while 
$\mu, \nu=0,\ldots,4$.

With such an ansatz the equations of motion (\ref{T0}--\ref{W0}) 
can be written as
\begin{eqnarray}
0 &=& R^{(4)}_{\al\beta}+\na_\al\pa_\beta \si
-g_{\al\beta}\left[e^{2A}\left(A''+4(A')^2-A'\phi'\right)\right] 
,
\label{EoMab}
\\[2pt]
0 &=& -4 A''-4(A')^2+\phi''\,,
\label{EoMyy}
\\[2pt]
0 &=& 2\La-\phi''-4A'\phi'+(\phi')^2
+e^{-2A}
\big[\pa^\al \si \pa_\al \si-\na^\al\pa_\al \si\big],
\label{EoMdil}
\end{eqnarray}
where prime denotes the derivative with respect to $y$.

For the trivial case $A(y)=\const$, $\phi'(y)=\const$ 
the eqs.\ (\ref{EoMab}--\ref{EoMdil}) just reduce 
to eqs.\ (\ref{T0}--\ref{W0}) with fields depending only on $x$ 
(with possibly different value of $\La$).

We now concentrate on the more interesting situation when 
$A(y)\ne\const$ and the metric (\ref{ds2}) is really a warped one. 
In such a case the coordinates $x$ and $y$ must separate in 
eq.\ (\ref{EoMab}) and we obtain two conditions:
\begin{eqnarray}
e^{2A}\left(A''+4(A')^2-A'\phi'\right)
&=&
\ga
\label{gamma}
\,,
\\[2pt]
R^{(4)}_{\al\beta}-\ga g_{\al\beta} +\na_\al\pa_\beta \si
&=&
0
\,.
\label{EoM4R}
\end{eqnarray}
Similarly the $x$--dependent part of eq.\ (\ref{EoMdil}) reads
\begin{equation}
\pa^\al \si \pa_\al \si-\na^\al\pa_\al \si
=
-2\la
\,.
\label{EoM4dil}
\end{equation}
However equations (\ref{EoMyy}), (\ref{EoMdil}) and (\ref{gamma})
are compatible with (\ref{EoM4dil})
only for vanishing constant $\la$. 
The 4--dimensional metric and dilaton must therefore satisfy 
eqs.\ (\ref{EoM4R}) and (\ref{EoM4dil}) with $\la=0$.
Acting on eq.\ (\ref{EoM4R}) with
$\na^\al$ and using eq.\ (\ref{EoM4dil}) gives the condition
\begin{equation}
\ga\pa_\beta \si=0
\,.
\end{equation}
Thus  either $\si=\const$ or $\ga=0$ and the
$x$--dependent parts of the warped solutions must 
satisfy one of the two following sets of equations:
\\
Case I
\begin{equation}
R^{(4)}_{\al\beta}=\frac{R}{4} g_{\al\beta}, \qquad R=\const,
\qquad \si=\const\,.
\label{EoM4I}
\end{equation}
Case II
\begin{equation}
R^{(4)}_{\al\beta}+\na_\al\pa_\beta \si=0\,,\qquad
\pa^\al\si\pa_\al\si=\na^\al\pa_\al\si\,,\qquad
\si\ne\const\,.
\label{EoM4II}
\end{equation}
The case I corresponds to the 4--dimensional equations of motion 
derived from the 4--dimensional action
\begin{equation}
S^I=
\frac1{2\ka^2}\int \d^4 x
\sg \left(-2\tilde\La+R^{(4)}\right),
\end{equation}
and the case II corresponds to the action
\begin{equation}
S^{II}=
\frac1{2\ka^2}\int \d^4 x
\sg e^{-\si}\left(R^{(4)}+\pa_\al\si\pa^\al\si\right).
\end{equation}
It is interesting to note that none of these effective actions is the
4--dimensional version of the general dilaton gravity action given in 
eq.\ (\ref{action}). From the effective 4--dimensional point of view 
the warped solutions cannot simultaneously have the 4--dimensional
cosmological constant and the nontrivial dilaton. This feature points
once again to the far--reaching differences between effective theories
obtained by dimensional reduction (where there may be additional
nontrivial constraints on the parameters) and theories defined ab
initio in 4 dimensions (where such constraints are absent).

\section{Exact solutions in the string frame}

It turns out that for both cases I and II eqs.\
(\ref{EoMab}--\ref{EoMdil}) can be solved for $A(y)$ and $\phi(y)$ 
in full generality.

Let us first discuss the case I with $R\ne0$ which from the
4--dimensional point of 
view has nonvanishing effective cosmological constant and
trivial 4--dimensional dilaton. The $y$-dependent part
of eqs.\ (\ref{EoMab}--\ref{EoMdil}) reads then 
\begin{eqnarray}
R&=&4e^{2A}\left(A''+4(A')^2-A'\phi'\right)\,,
\label{caseI1}
\\[2pt]
0 &=& -4 A''-4(A')^2+\phi''\,,
\label{caseI2}
\\[2pt]
0 &=& 2\La-\phi''-4A'\phi'+(\phi')^2\,.
\label{caseI3}
\end{eqnarray}
These equations are dependent -- the derivative of (\ref{caseI1}) 
combined with (\ref{caseI2}) and its derivative gives the derivative of 
(\ref{caseI3}). 

The character of the solutions
depends on the sign of the bulk cosmological constant $\La$. For
$\La>0$ we choose the integration constants ($\al$, $y_1$, $\phi_0$)
in such a way that the most general solution can be written in the
following form 
\begin{eqnarray}
e^{2A(y)}&=&\frac{R}{2\La}
+\frac{R}{2\La}\cot\left({\sqrt{\La/2}}(y-y_1)\right)
\left(\pi\al-
\sqrt{\La/2}(y-y_1)\right),
\label{ALp}
\\
\phi(y)&=&\phi_0+\ln\left[
\frac{R}{2\La}
+\frac{R}{2\La}\cot\left(\sqrt{\La/2}(y-y_1)\right)
\left(\pi\al-
\sqrt{\La/2}(y-y_1)\right)\right]\nn\\
&&-\ln\left(\sin^2\left(\sqrt{\La/2}(y-y_1)\right)\right),
\label{phiLp}
\end{eqnarray}
where the allowed values of the constant $\al$ are $\al>0$ or
$\al\le-1$. Formally the above solutions have infinitely many branches
separated by singularities of $A(y)$ and $\phi(y)$. But it is enough to
consider only one branch for a given set of the integration
constants. That special branch is defined as the one that starts at
$y_1$ and for which $A(y)\to+\infty$ when $y\to y_1$. All other
branches can be obtained, up to a shift in $y$, by some other choice
of the integration constants.

The special branch connected to $y_1$ has also
a second singularity of $A(y)$. For $0<\al<1$ the warp function
$A(y)$ goes at that singularity to $+\infty$ while for $|\al|\ge 1$ 
it goes to $-\infty$. The solutions with $A(y)\to+\infty$ at both
singularities exist only when the effective 4--dimensional curvature
scalar $R$ is positive. In such a case we take the branch
$y_1<y<y_1+\pi\sqrt{2/\La}$. The solutions with $A(y)\to-\infty$ at
one of the singularities can be obtained for both signs of $R$. The
range of $y$ in such a case is between $y_1$ and the next singularity.
We take the singularity to the right (left) from $y_1$ when the
product  $\al R$ is positive (negative). The position of that
singularity can not be found in a closed form. It is a value of $y$
for which the right hand side of eq.\ (\ref{ALp}) vanishes.

We now turn to the situation when the bulk cosmological constant 
$\La$ is negative. For this case we introduce three integration
constants ($y_0$, $y_1$, $\phi_0$) and the additional discrete
parameter $\ka=\pm1$. The solutions for $\La<0$ read 
\begin{eqnarray}
e^{2A(y)}&=&\frac{R}{2\La}
-\frac{R}{2\La}
\coth{^\ka}\left({\sqrt{-\La/2}}(y-y_1)\right)\cdot
\nn\\
&&
\qquad\qquad\cdot\left(\sqrt{-\La/2}(y-y_0)
+\tanh{^\ka}\left({\sqrt{-\La/2}}(y_0-y_1)\right)\right),
\nn\\ 
\label{ALn}\\
\phi(y)&=&\phi_0+\ln\left[
\frac{R}{2\La}
-\frac{R}{2\La}
\coth{^\ka}\left(\sqrt{-\La/2}(y-y_1)\right)\cdot\right.
\nn\\
&&
\left.\ \qquad\qquad\cdot\left(\sqrt{-\La/2}(y-y_0)
+\tanh{^\ka}\left(\sqrt{-\La/2}(y_0-y_1)\right)
\right)\right]\nn\\
&&\ 
-\ln\left(\frac{1-\ka}{2}+\sinh^2(\sqrt{-\La/2}(y-y_1))\right).
\label{phiLn}
\end{eqnarray}
In this case there is no problem of multiple branches. For $\ka=+1$
the real solutions exist for $y$ inside the interval bounded by
$y_0$ and $y_1$ for $R<0$ and outside this interval for $R>0$. At each
of the points, $y_0$ and $y_1$, there is a singularity of $A(y)$ (and
also of $\phi(y)$) but the nature of these singularities is different:
$A(y)\to-\infty$ for $y\to y_0$ while $A(y)\to+\infty$ for $y\to y_1$.

The solutions for $\ka=-1$ are defined for $y$ inside the interval
bounded by $y_0$ and $y_2$ for $R<0$ and outside this interval  
for $R>0$. The boundary position $y_2$ is defined as the second
(besides $y_0$) zero of the right hand side of eq.\ (\ref{ALn}). 
In this case it is different from $y_1$ and cannot be
written in a simple form. At both points, $y_0$ and $y_2$, the warp
function $A(y)$ approaches $-\infty$.

The solution for $\La=0$ can be obtained as a limit of (\ref{ALp}) 
and (\ref{phiLp}):   
\begin{eqnarray}   
e^{2A(y)}&=&\frac{R}{12}\frac   
{(y-y_1)^3-(y_0-y_1)^3}{y-y_1},
\label{AL0}
\\[2pt]
\phi(y)&=&\ln\left[\frac{C_\phi R}{12}\frac
{(y-y_1)^3-(y_0-y_1)^3}{(y-y_1)^3}\right],
\label{phiL0}
\end{eqnarray}
where we changed the integration constants in order to
simplify the result. The above solutions have singularities at $y_0$
and $y_1$ such that $A(y)\to-\infty$ for $y\to y_0$ and 
$A(y)\to+\infty$ for $y\to y_1$. For $R<0$ ($R>0$) the solution is
well defined for $y$ inside (outside) the interval bounded by $y_0$
and $y_1$.

Let us now discuss case I with $R=0$ and case II (they lead to the
same $y$-dependent equations). The equations we have to solve are 
given by (\ref{caseI1}--\ref{caseI3}) with $R=0$.

The solutions for $\La>0$ read:
\begin{eqnarray}   
e^{2A(y)}&=&c_A\left|\tan\left(\sqrt{\La/2}(y-y_0)\right)
\right|,
\label{ALpII}
\\[2pt]
\phi(y)&=&\phi_0+2\ln\left|\tan\left(\sqrt{\La/2}(y-y_0)\right)
\right|
-\ln\left|\sin\left(\sqrt{2\La}(y-y_0)\right)\right|.
\label{phiLpII}
\end{eqnarray}
The range of $y$ is  either $y_0<y<y_0+\pi\sqrt{2/\La}$ or 
$y_0-\pi\sqrt{2/\La}<y<y_0$. The warp function $A(y)$ goes to
$-\infty$ at $y=y_0$ and to $+\infty$ at the second singularity.

In the case $\La<0$ we again introduce additional parameter
$\ka=\pm 1$ and write the solutions in the form:
\begin{eqnarray}
e^{2A(y)}&=&c_A\left|
\tanh\left(\sqrt{-\Lambda/2}(y-y_0)\right)
\right|^\ka,
\label{ALnII}
\\[2pt]
\phi(y)&=&\phi_0+2\ka\ln\left|
\tanh\left(\sqrt{-\La/2}(y-y_0)\right)\right|
-\ln\left|\sinh\left(\sqrt{-2\La}(y-y_0)\right)\right|,
\label{phiLnII}
\end{eqnarray}
where $y>y_0$ or $y<y_0$.

Finally for $\La=0$ we obtain:
\begin{eqnarray}
e^{2A(y)}&=&c_A\left|y-y_0\right|^\ka,
\label{AL0II}
\\[2pt]
\phi(y)&=&(2\ka-1)\ln \left(C_\phi\left|y-y_0\right|\right),
\label{phiL0II}
\end{eqnarray}
where again $y>y_0$ or $y<y_0$. In the last two types of solutions
(\ref{ALnII}--\ref{phiL0II}) we have $\ka A(y)\to -\infty$ for 
$y\to y_0$.

Let us now summarize the structure of singularities for all the
solutions. We denote by $S_+$ (respectively $S_-$) singularity at
finite $y$ where $A(y)\to +\infty$ (respectively $-\infty$) and by
$S_{inf}$ the fact that a solution extends to $y\to\infty$ or
$y\to-\infty$. The character of the solutions depends on the signs of
the bulk cosmological constant $\La$ and the effective 4--dimensional
curvature scalar $R$ and also on the presence of a nontrivial
4--dimensional dilaton (case II). All the cases are summarized in the
following table: 
\begin{center}
\begin{tabular}{|c|c|c|c|}
\hline
&\ case I, $R<0\ $&\ case I, $R=0\ $&\ case I, $R>0\ $ \\ 
&&and case II&\\[2pt]  
\hline
$\,\,\La>0\,\,$ & $\,\,S_+-S_-\,\,$ & $\,\,S_+-S_-\,\,$ 
& $\,\,S_+-S_+(S_-)\,\,$    
\\[2pt] 
\hline
$\,\,\La=0\,\,$ & $\,\,S_+-S_-\,\,$ & $\,\,S_+(S_-)-S_{inf}\,\,$ 
& $\,\,S_+(S_-)-S_{inf}\,\,\,\,$  
\\[2pt]
\hline
$\,\,\La<0\,\,$ & $\,\,S_--S_+(S_-)\,\,$ & $\,\,S_+(S_-)-S_{inf}\,\,$ 
& $\,\,S_+(S_-)-S_{inf}\,\,\,\,$  
\\[2pt]
\hline
\end{tabular}
\end{center}

One can see that each solution has at least one singularity at 
finite $y$. Solutions with non--positive bulk cosmological constant
$\La$ and non--negative effective curvature scalar (case I with
$R\ge0$ and case II) start at such singularity and extend to
infinity. All other solutions have two singularities at finite values
of $y$. The distance between the singularities is of the order of
$1/\sqrt{\La}$ for $\La>0$ and is a free parameter for $\La\le 0$.

\section{Solutions in the Einstein frame}

So far we have presented general warped solutions in 5--dimensional
dilaton gravity in the string frame. Let us now
discuss the form of these solutions in the Einstein frame.

In order to go to the Einstein frame we perform a Weyl transformation
for which $\vphi(x,y)$ is unchanged and the metric transforms as
\begin{equation}
\tg_{\mu\nu}=e^{-2\vphi/3} g_{\mu\nu}
\,.
\end{equation}
Then the 5--dimensional action reads
\begin{equation}
S_E=\frac1{2\ka^2}\int \d^5 x
\sqrt{-\tg}\left(-2\La e^{\frac23\vphi}+\tilde{R}^{(5)}
-\frac13\pa_\mu\vphi\pa^\mu\vphi\right).
\label{actionE}
\end{equation}
Of course we are not going to solve the equations of motion derived
from the above action since we can simply translate the solutions
obtained before in the string frame. These solutions have the warped
form also in the Einstein frame only for $\si=\const$ (i.e. case I). 
Then we have
\begin{equation}
ds^2_E=e^{2A_E(y)}\tilde g_{\al\beta}(x)dx^\al 
dx^\beta+ e^{-2\phi(y)/3}dy^2,
\label{ds2E}
\end{equation}
where
\begin{eqnarray}
e^{2A_E(y)}
&=&
e^{2A(y)-2\phi(y)/3}\,,
\label{AEphiE}
\end{eqnarray}
In order to put the metric (\ref{ds2E}) in the standard form
%corresponding to vanishing $B_E$ 
one should redefine the last
coordinate by the relation
\begin{equation}
dz=e^{-\phi(y)/3}dy\,.
\label{zytrans}
\end{equation}
Using this new coordinate $z$ the line element in the Einstein frame
can be written in the usually assumed form
\begin{equation}
ds^2_E
=
e^{2\tilde{A}_E(z)}\tilde g_{\al\beta}(x)dx^\al dx^\beta+ dz^2
\,.
\label{ds2Ez}
\end{equation}
However, in general there is no analytic formula for
$\tilde{A}_E(z)$ because for the string--frame general solutions 
found in the paper it is not possible to integrate and invert in
closed form the equation (\ref{zytrans}).
One can transform our general
string--frame solutions to the Einstein frame either
as explicit solutions for a modified form of the metric
(\ref{ds2E}) using the coordinate $y$ or as non-explicit solutions for
a standard form of the metric (\ref{ds2Ez}). Although the latter
transformation is quite complicated but 
nevertheless some important general remarks can be made. The most
interesting one is related to the character of singularities at finite
$z$. It is not very difficult to check that for all the solutions
both $S_+$ and $S_-$ singularities in the string frame are 
translated into $S_-$ singularity (i.e. $\tilde{A}_E(z)\to -\infty$)
in the Einstein frame. One can show also that finite distance between
singularities in the string frame in the $y$ coordinate is also a
finite distance between singularities in the $z$ coordinate. 

These results allow us to divide the solutions into two classes. 
In one class the solutions end at two points at which the metric
vanishes and some components of the curvature tensor have 
singularities. The distance between these points is finite. 
The solutions have such features when $\La>0$ or $R<0$. For other
parameters, namely when $\La\le0$ and $R\ge0$, a solution 
has only one singularity and extends infinitely in one of the
directions in the coordinate $y$ (the same is true also for coordinate 
$z$). The metric vanishes at the singularity and grows to infinity for
$|y|\to\infty$ ($|z|\to\infty$).

Any of the globally well defined warped solutions in such a theory has
to be piecewise (i.e. outside the branes) given by the solutions found
in this paper since as we have shown these are the most general
warped solutions of the bulk equations of motion.
If branes with appropriate properties are added then we can eventually
build models which are effectively 4--dimensional.

We now discuss situations when it is possible to find
explicitly the solutions in terms of the coordinate $z$. They
correspond to case I ($\si=\const$) with vanishing 4--dimensional
curvature scalar $R$. For $\La=0$, using eqs.\ (\ref{AL0II}),
(\ref{phiL0II}) and (\ref{zytrans}) we get
\begin{eqnarray}
e^{2\tilde A_E(z)}
&=&
\sqrt{\tilde C_A\left|z-z_0\right|}\,,
\\[2pt]
\phi(z)
&=&
\frac{3}{2}
\ln\left(\tilde C_\phi\left|z-z_0\right|^\ka\right).
\end{eqnarray}
where the new coordinate $z$ is given by $z=z_0
+C_z(y-y_0)^{2\ka/(2\ka+1)}$.
These are the solutions discussed in ref.\ \cite{KSS1} (with different 
normalization of the dilaton). For nonzero bulk cosmological constant
$\La$ we obtain
\begin{eqnarray}
e^{2\tilde A_E(z)}
&=&
\sqrt{\tilde C_A
\left(z-z_0\right)
\left[\left(z_1-z_0\right)^3-\left(z-z_0\right)^3\right]}
,\,\,\,\,\label{metrELn}
\\[2pt]
\phi(z)
&=&
\frac32
\ln\left[\frac{2}{9\La}
\frac{z-z_0}{\left(z_1-z_0\right)^3-\left(z-z_0\right)^3}\right].
\label{phiELn}
\end{eqnarray}
For $\La>0$ ($\La<0$) the
solution is well defined for $z$ inside (outside) the interval bounded
by $z_0$ and $z_1$, at which the metric vanishes. 
The coordinate $z$ for $\La>0$ is equal to 
\begin{equation}
z=z_0+
C_z\left[\sin\left(\sqrt{\La/2}(y-y_0)\right)\right]^{2/3},
\end{equation}
while for $\La<0$
\begin{eqnarray}
z&=&z_0+
C_z\left[\sinh\left(\sqrt{\La/2}(y-y_0)\right)\right]^{2/3},
\\ 
z&=&z_1+
C_z\left[\cosh\left(\sqrt{\La/2}(y-y_0)\right)\right]^{2/3},
\end{eqnarray}
for $\ka=+1$ and $\ka=-1$ respectively.

The solutions in the Einstein frame (\ref{metrELn}--\ref{phiELn}) are
new. Let us emphasize that these solutions constitute only a
small subset of general solutions found in the paper (the others in
general cannot be explicitly transformed to the Einstein frame and
only some partial results can be obtained for example for $R\ne 0$ and 
$\La=0$ \cite{KSS2}).

\section{Conclusions}

The general warped solutions of the 5--dimensional
dilaton gravity presented in this paper may be useful in the
brane--like applications of string inspired models and supplemented by
appropriate branes may lead to globally well defined warped solutions. 
It is worth emphasizing once again that the 5--dimensional theory
admits warped solutions only of two 4--dimensional types: Einstein
metric with a constant dilaton or nontrivial dilaton without a
cosmological constant. 
This observation may have important
consequences for all theories that include 4--dimensional dynamical 
dilaton.

The solutions described in this paper have curvature singularities (in
the 5--di\-men\-sional 
sense). It is therefore necessary to supplement them with appropriate
branes to obtain solutions which are well behaved for the whole range
of $y$. It is easy to transpose the solutions to the ``cosmological
case'' where the distinguished coordinate is not $y$ but $t$ which is
timelike.  It is then not necessary to add
any branes because the evolution starts
(and eventually ends) at singularities which are of the same type as
in standard cosmology. 

\vspace{5mm}
\noindent
{\Large\bf Acknowledgements}
\vspace{5mm}

Work supported in part by the European Community's Human Potential
Programme under contract
HPRN--CT--2000--00152 Supersymmetry and the Early Universe
and by the Polish KBN grant 2 P03B 001 25.

\end{document}